# Non-invasive imaging through thin scattering layers with broadband illumination

TENGFEI WU,[1] CHENGFEI GUO,[1] XIAOPENG SHAO[1,*]

[1]School of Physics and Optoelectronic Engineering, Xidian University, Shaanxi 710071, China
*Corresponding author: xpshao@xidian.edu.cn

**Memory-effect-based methods have been demonstrated to be feasible to observe hidden objects through thin scattering layers, even from a single-shot speckle pattern. However, most of the existing methods are performed with narrowband illumination or require point light-sources adjacent to the hidden objects as the references, to make an invasive pre-calibration of the imaging system. Here, inspired by the shift-and-add algorithm, we propose that by randomly selecting and averaging different sub-regions of the speckle patterns, an image pattern resembling the autocorrelation (we call it R-autocorrelation) of the hidden object can be extracted. By performing numerical simulations and experiments, we demonstrate that comparing with true autocorrelation, the pattern of R-autocorrelation has a significantly lower background and higher contrast, which enables better reconstructions of hidden objects, especially in the case of broadband illumination, or even with white-light.**

Due to the light scattering caused by the inhomogeneity inside the scattering media, such as tissues, any object can hardly be observed with a conventional optical imaging system, only a complex speckle pattern is obtained instead [1]. Interestingly, the speckle pattern contains much information of the hidden objects and in recent years many significant methods have been proposed to retrieve the image of the hidden objects. Wavefront shaping is one of the earliest techniques to solve this intractable problem and it has been demonstrated to realize focusing or imaging through scattering media [2-10]. However, this method requires a detector or a probe in the plane of interest, which is always considered to be avoided in real applications. Moreover, a complex and time-consuming calibration process makes it difficult to be used for dynamic samples. Holographic imaging has also been demonstrated as a possible way [11]. However, a reference is always required during the imaging process.

Since a recent breakthrough, i.e. speckle-correlation imaging method, was proposed by Bertolotti et al. [12], memory-effect, which states the inherent angular correlation of scattered light, has been considered as one of potential phenomena to be exploited to realize non-invasive imaging through thin scattering layers [13-14]. Based on the same concept and inspired by the astronomy imaging, Katz et al. demonstrated that by calculating the autocorrelation of a single-shot high-resolution speckle pattern, the Fourier amplitude of the hidden object can be extracted [15]. A phase-retrieval algorithm is then used to recover the object image [16]. In the following years, several memory-effect-based methods were proposed to extract more useful information from the speckle patterns [17-19]. However, most of these methods are demonstrated with a narrowband illumination source. Although the broadband illumination can fundamentally be used in some methods, it may significantly increase the background and reduce the contrast of the obtained object autocorrelation, which can wash out some object features [15, 20]. Deconvolution is also a potential method to reconstruct the speckle pattern without losing the information, even with the white-light illumination [21-22]. However, it always requires one or several light point-sources adjacent to the hidden objects to make an invasive pre-calibration of the whole imaging system.

In this Letter, inspired by the shift-and-add algorithm used in the astronomy imaging [23], we propose a method to realize non-invasive imaging through thin scattering layers with broadband illumination. We randomly select a series of different sub-regions of the speckle patterns, and ensure the intensity value of each center pixel of the sub-regions is the maximum or a near maximum. By averaging all the sub-regions, we can obtain an image pattern, which we call R-autocorrelation, resembling the true autocorrelation of the hidden objects and containing the object information. By performing the numerical simulations and the experiments, we demonstrate that the pattern of R-autocorrelation possesses a lower background and higher contrast than the true autocorrelation, especially in the case of broadband illumination, which determines the object images can be recovered with clearer structures by using R-autocorrelation.

The concept of non-invasive imaging with R-autocorrelation is presented in Fig. 1. An object (Fig. 1(a)) is hidden behind a scattering layer. If we assume that it lies within the range determined by the memory-effect and is illuminated by a narrowband illumination source, the system can be considered as an incoherent imaging system with a shift-invariant point-spread function (PSF) [15]. The camera image $I$ (Fig. (c)) is a convolution of the object $O$ and the PSF of the system (Fig. 1(b)), and it can be expressed as follows without considering the magnification:

$$I = O * PSF \qquad (1)$$

where "$*$" denotes the convolution operator. Fig. 1(d) shows a series of sub-regions randomly selected from a high-resolution camera image (see below for more details), and R-autocorrelation can then be obtained by averaging the $N$ sub-regions. Finally, a phase-retrieval algorithm is used, combining with the Fourier amplitude information of the hidden object extracted from the R-autocorrelation, to retrieve the object image (Fig. 1(f)).

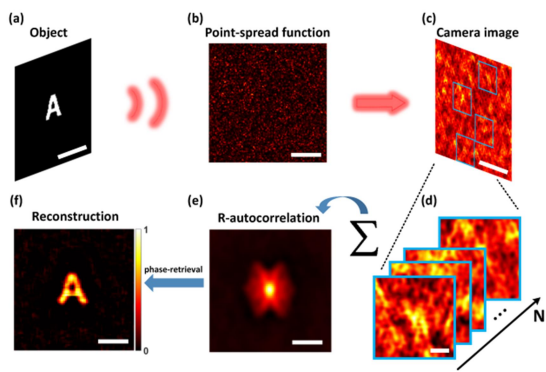

Fig. 1. Concept of non-invasive imaging through thin scattering layers with R-autocorrelation in the case of narrowband illumination. (a) True object; (b) point-spread function of the scattering imaging system; (c) speckle pattern; (d) different sub-regions of the speckle pattern; (e) recovered R-autocorrelation of the object by averaging the sub-regions; (f) reconstruction from the R-autocorrelation. Scale bar: 20 pixels in (a), (d), (e) and (f) and 600pixels in (b) and (c).

Shift-and-add algorithm is one of earliest methods used in the astronomy imaging to reduce the influence of the atmospheric turbulence and improve the imaging quality of telescopes [23]. Assuming that an object is composed by several point-sources, in which one of them possesses the dominant intensity. Therefore, the brightest part in the degraded image should also be the brightest part in the true image. Given a series of degraded images at different moments, one can then shift the brightest part to the center of each image, and the average of all the degraded images can recover an object image with higher quality. The whole process can be simply written as [23]:

$$R(x, y) = \sum_{m=1}^{M} S_m(x - x_m, y - y_m) \qquad (2)$$

where $M$ is the frame number. $x$ and $y$ are the coordinates in the imaging plane, $x_m$ and $y_m$ denote the coordinates of the brightest point in $m$th degraded image $S_m$, and $R$ is the result obtained by the summation of all the shifted degraded-images.

In the highly scattering cases, a speckle pattern can be considered as the random superposition of many replicas of object images generated at the position of each speckle grain. Considering the autocorrelation, any sub-region, which is larger than twice of the size of the object image, should contain the complete structure information of the object. Therefore, the idea in the shift-and-add algorithm can be interestingly realized in the highly scattering cases by dividing the high-resolution speckle pattern into many random sub-regions. In our realization, we first randomly select several initial sub-regions on the camera image; secondly, the position of the pixel with the highest intensity value in each sub-region is recorded and the updated sub-regions are then re-selected on the speckle pattern with the recorded positions as their centers, as is shown in Fig. 1(d). The intensity value of center pixels in the updated sub-regions should be the maximum or a near-maximum. It is worth noting that the object image cannot be simply retrieved with the aforementioned operation, since two important factors determine that we can hardly locate the same part of the object in the center of each sub-region. Firstly, in most applications the hidden objects have a certain intensity distribution without a dominant point-source, and secondly, the random overlaps of the image replicas on different speckle grains also have an inevitable influence. However, it allows us to obtain an image pattern, which resembles the true autocorrelation of the hidden object, and we call it R-autocorrelation (Fig. 1(e)) [24-25]. As in the speckle-correlation methods, the Fourier amplitude of the object can be extracted from the R-autocorrelation and the reconstruction of the object image is obtained with a phase-retrieval algorithm (Fig. 1(f)).

To compare the performance of R-autocorrelation and true autocorrelation, some numerical simulations are performed. We use the Fresnel diffraction to describe the light propagation and there are 80 pixels in each dimension of a sub-region. When the illumination source has a certain bandwidth, the PSF is no longer fixed, and instead, it can be described as a function of wavelength $PSF(\lambda)$, where $\lambda$ denotes the wavelength. The camera image $I_B$ is actually a superposition of the speckle patterns generated by different wavelengths:

$$I_B = \sum_{\lambda} \alpha_\lambda \cdot [O * PSF(\lambda)] \qquad (3)$$

where $\alpha_\lambda$ is the weight coefficient, representing the ratio of each wavelength in the light source or the wavelength response of the camera.

In the narrowband case, an illumination source with a bandwidth of ~1nm, centered at 632.8nm is performed. A Gaussian function is introduced to work as $\alpha_\lambda$. We calculate the R-autocorrelation of the speckle pattern with the aforementioned operations and show the result in the first row of Fig. 2(a). The true autocorrelation is shown in the first row of Fig. 2(b). Fig. 2(c) shows the comparison of one column, marked by the green (R-autocorrelation) and the red (true autocorrelation) dashed lines, of the two image patterns. The results indicate that the R-autocorrelation has a sharper structure and higher contrast, compared to the true autocorrelation. The reconstructions (the second row in Fig. 2(a) and Fig. 2(b)) are obtained by using the basic phase-retrieval algorithm, i.e. the combination of Hybrid Input-Output (HIO) and Error Reduction (ER) algorithms [12, 15, 16]. Fig. 2(d) shows the true object.

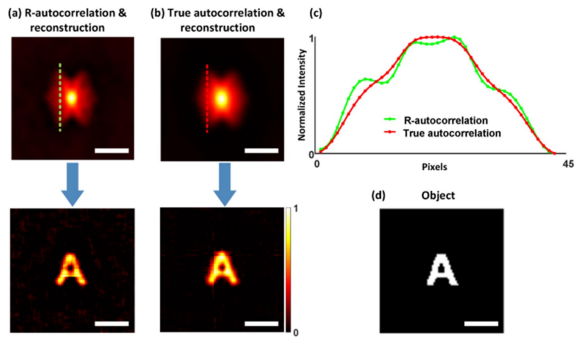

Fig. 2. Numerical simulations of imaging with R-autocorrelation and true autocorrelation in the case of narrowband illumination. (a) R-autocorrelation and corresponding reconstruction; (b) true autocorrelation and corresponding reconstruction; (c) dashed lines in (a) and (b) (green and red); (d) true object. Scale bar: 20 pixels.

In the case of broadband illumination, the speckle pattern is generated by the simultaneous contribution from different wavelengths, as is described in Eq. (3). When the wavelengths are too far apart, an "exploding speckle pattern" is formed and most of the speckle grains are smeared out and lose the information, compared to the narrowband illumination [23, 26]. However, in the central part of the speckle pattern, some speckle grains have a significant probability to overlap and keep the contrast, which allows us to extract the object information. Fig. 3(a) is the central part (600×600 pixels) of a speckle pattern that is generated by a broadband illumination. The wavelength is from 500nm to 764nm with a sampling of 0.5nm. A Gaussian function is used to weight each wavelength and the full width at half maximum (FWHM) is around 104nm. The speckle pattern in the central part contains the required object information, although the speckle contrast is obviously lower than the narrowband case (Fig. 3(b)). Since most of the speckle grains are smeared out with the broadband illumination, more frames are always required to sufficiently suppress the statistical noise. The R-autocorrelation and the true autocorrelation obtained with 30 frames in the case of broadband illumination are respectively shown in the first row in Fig. 3(c) and Fig. 3(d), in which we notice that the R-autocorrelation can significantly suppress the strong background appeared in the true autocorrelation and it has a higher contrast of the object information. The reconstruction from the R-autocorrelation also has a clearer structure than that from the true autocorrelation, as are shown in the second row.

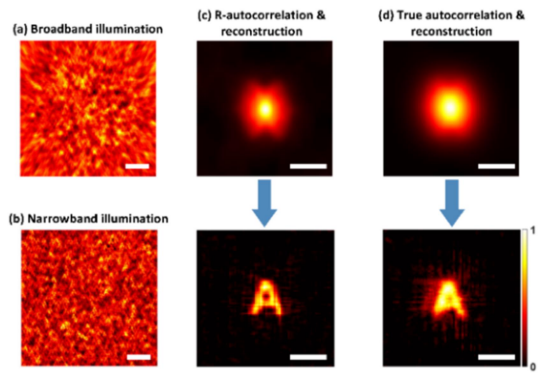

Fig. 3. Numerical simulation of imaging with R-autocorrelation in the case of broadband illumination. (a) Central part of speckle pattern with broadband illumination; (b) central part of speckle pattern with narrowband illumination; (c) and (d) are the R-autocorrelation, the true autocorrelation, and the corresponding reconstructions in the case of broadband illumination. Scale bar: 100 pixels in (a) and (b), 20 pixels in (c) and (d).

Fig. 4 is the optical setup of the experimental demonstration of this concept. The light-source is a quartz tungsten-halogen lamp (Thorlabs, QTH10). The incoherent light illuminates the object and then reaches the position of a thin scattering layer (Edmund, Ground Glass Diffuser), which is placed ~60cm away from the object. An iris with a diameter of ~3.3mm is placed against the scattering layer, to control the spatial resolution of the imaging system. The speckle patterns are captured by a high-performance camera (Andor, ZYLA-5.5-USB3.0, 2160×2560), which is placed ~12cm in front of the scattering layer. Considering the real spectrum of the light-source and the wavelength response of the camera, the FWHM of the effective spectrum of the illumination source should be around 220nm.

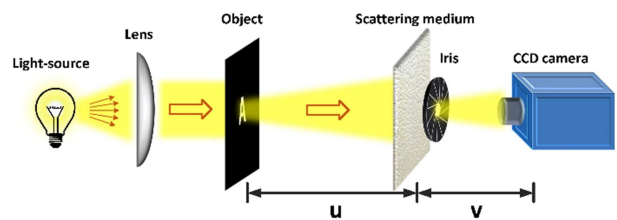

Fig. 4. Experimental setup of non-invasive imaging with broadband illumination.

The objects used in the experiments are printed in the transparencies, and each of them is around 700um. Since most of the speckle grains in the speckle pattern are smeared out in the case of broadband illumination, we select the central part (600×600 pixels) of the camera images to retrieve the hidden objects. By moving the thin scattering diffuser, 40 frames of uncorrelated speckle patterns, for recovering each object, are collected to sufficiently suppress the statistical noise when calculating the true autocorrelation or the R-autocorrelation. A sub-region with 80 pixels in each dimension is selected and we use $10^5$ sub-regions in each frame to calculate the R-autocorrelation of the hidden objects. To experimentally demonstrate that calculating the R-autocorrelation can suppress the strong background of the true autocorrelation in the case of broadband illumination, we compare the R-autocorrelation and the true autocorrelation of the object "letter A" in Fig. 5(a) and Fig. 5(b), as an example. Fig. 5(c) is the calculated true autocorrelation of the object "letter A" from a single-shot narrowband speckle pattern, which is generated with a narrowband illumination source, i.e. the combination of a spatially incoherent light-emitting diode (Thorlabs, M625L3) and a narrow band-pass filter (Andover, 633FS02-50, 1.0+/−0.2nm). It is interesting to notice that in our case, the R-autocorrelation obtained with a broadband illumination is even much similar as the true autocorrelation obtained in the case of narrowband illumination. The embedded images are the true object "letter A" used in the experiments.

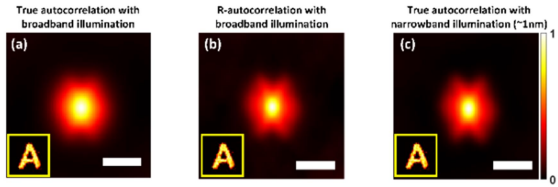

Fig. 5. Comparison of true autocorrelation and R-autocorrelation. (a) True autocorrelation with broadband illumination; (b) R-autocorrelation with broadband illumination; (c) true autocorrelation with narrowband illumination. Scale bar: 20 pixels.

Fig. 6(a) shows some of the 40 acquired frames of the raw camera images for different objects. For recovering the image, we use the basic phase-retrieval algorithm. We run 5100 iterations of HIO, in which the factor $\beta$ gradually decreases from 2 to 0 with a step of 0.04. The result of HIO is then used as an initial guess of ER to run another 100 iterations to complete the reconstruction. To realize a faithful reconstruction of each image, 200 independent runs (50 runs for numerical simulations) are performed with different random initial guesses and the reconstruction closest to the true object is selected as the final reconstruction. Fig. 6(b) and Fig. 6(c) compare the reconstructions with true autocorrelation and R-autocorrelation. The retrieved images from the R-autocorrelation are more recognizable, since they have clearer structures and lower backgrounds. The R-autocorrelations and the true autocorrelations for each reconstruction are also embedded in Fig. 6(b) and Fig. 6(c). The true objects are shown in Fig. 6(d) as references.

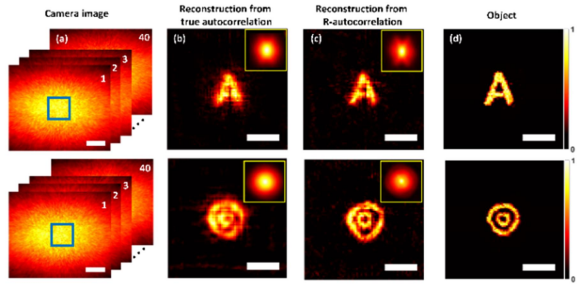

Fig. 6. Experimental results. (a) Camera images. The selected central part is marked by the solid box; (b) reconstructions from true autocorrelation; (c) reconstructions from R-autocorrelation. Embedded images: true autocorrelations and R-autocorrelations; (d) true objects. Scale bar: 500 pixels in (a), 20 pixels in (b), (c) and (d).

In conclusion, we propose a non-invasive imaging method to observe the objects hidden behind thin scattering layers with broadband illumination. A large amount of sub-regions are randomly selected on the speckle patterns and the center pixel of each sub-region should possess the maximum or near maximum intensity value. By averaging all the sub-regions, the R-autocorrelation, resembling the true autocorrelation of objects, can be obtained, thanks to the shift-and-add algorithm. By performing the numerical simulations and experiments, we demonstrate the R-autocorrelation contains the object information, and a phase-retrieval algorithm can be used to recover the object image from the R-autocorrelation. Furthermore, in the case of broadband illumination, calculating the R-autocorrelation can significantly suppress the pattern background and enables better reconstructions, compared to the true autocorrelation. In addition, to our knowledge it is also the first experimental demonstration of the memory-effect based non-invasive imaging through thin scattering layers with white-light illumination (a quartz tungsten-halogen lamp). Compared to the deconvolution methods, we require no references to calibrate the imaging system. To this end, we freely make available the source codes and experimental data to use by the scientific community, as shown in the Codes and Data file.

**Funding.** National Natural Science Foundation of China (NSFC) (61575154); Fundamental Research Funds for the Central Universities (SA-ZD160501).

**Acknowledgment**. The authors would like to thank Xueen Wang and Ke Yin for the insightful discussions, and Lei Zhu for the experiments of measuring the spectrum of light source.